\documentclass[useAMS,aastex]{emulateapj}
\usepackage{longtable}

\def\approxgt{\ifmmode \rlap{$>$}{}_{{}_{{}_{\textstyle\sim}}} \else%
$\rlap{$>$}{}_{{}_{{}_{\textstyle\sim}}}$\fi} 
\def\approxlt{\ifmmode \rlap{$<$}{}_{{}_{{}_{\textstyle\sim}}} \else%
$\rlap{$<$}{}_{{}_{{}_{\textstyle\sim}}}$\fi}

\def\degr{\hbox{$^\circ$}}
\def\arcmin{\hbox{$^\prime$}}
\def\arcsec{\hbox{$^{\prime\prime}$}}

\def\flx{erg cm$^{-2}$ s$^{-1}$}
\def\lum{erg s$^{-1}$}

\def\chan{{\it Chandra}}

\def\i{$i^\prime$}
\def\r{$r^\prime$}

\shorttitle{The Galactic Bulge Survey}
\shortauthors{Jonker et al.}

\begin{document}

\title{The Galactic Bulge Survey: outline and X--ray observations}

\author{P.G.~Jonker\altaffilmark{1,2,3}} 

\email{p.jonker@sron.nl}
\author{C.G.~Bassa\altaffilmark{4,1,2}}
\author{G.~Nelemans\altaffilmark{2}}
\author{D.~Steeghs\altaffilmark{5,3}}
\author{M.A.P.~Torres\altaffilmark{1,3}}
\author{T.J.~Maccarone\altaffilmark{6}}
\author{R.I.~Hynes\altaffilmark{7}}
\author{S.~Greiss\altaffilmark{5}}
\author{ J.~Clem\altaffilmark{7}}
\author{A.~Dieball\altaffilmark{6}}
\author{V.J.~Mikles\altaffilmark{7}}
\author{C.T.~Britt\altaffilmark{7}}
\author{L.~Gossen\altaffilmark{7}}
\author{A.C.~Collazzi\altaffilmark{7}}
\author{R.~Wijnands\altaffilmark{8}}
\author{J.J.M.~In't Zand\altaffilmark{1}}
\author{M.~M\'endez\altaffilmark{9}}
\author{N.~Rea\altaffilmark{10}}
\author{E.~Kuulkers\altaffilmark{11}}
\author{E.M.~Ratti\altaffilmark{1}}
\author{L.M.~van Haaften\altaffilmark{2}}
\author{C.~Heinke\altaffilmark{12}}
\author{F. \"Ozel\altaffilmark{13}}
\author{P.J.~Groot\altaffilmark{2}}
\author{F.~Verbunt\altaffilmark{14,1}}

\altaffiltext{1}{SRON, Netherlands Institute for Space Research, Sorbonnelaan 2,
  3584~CA, Utrecht, The Netherlands}
\altaffiltext{2}{Department of Astrophysics, IMAPP, Radboud University Nijmegen,
Heyendaalseweg 135, 6525 AJ, Nijmegen, The Netherlands}
\altaffiltext{3}{Harvard--Smithsonian  Center for Astrophysics, 60 Garden Street, Cambridge, MA~02138, U.S.A.}
\altaffiltext{4}{Jodrell Bank Centre for Astrophysics, School of Physics and Astronomy, University of Manchester, Manchester M13 9PL, United Kingdom}
\altaffiltext{5}{Astronomy and Astrophysics, Department of Physics, University of Warwick, Coventry, CV4~7AL, United Kingdom}
\altaffiltext{6}{School of Physics and Astronomy, University of Southampton, Southampton SO17 1BJ, United Kingdom}
\altaffiltext{7}{Louisiana State University, Department of Physics and Astronomy,
Baton Rouge LA 70803-4001, U.S.A.}
\altaffiltext{8}{Astronomical Institute ``Anton Pannekoek'', University of Amsterdam, Kruislaan 403, 1098 SJ Amsterdam, The Netherlands}
\altaffiltext{9}{Kapteyn Astronomical Institute, University of Groningen, P.O.~Box 800, 9700 AV Groningen, The Netherlands}
\altaffiltext{10}{Institut de Ciencies de l'Espai (ICE, IEEC-CSIC), Campus UAB, Fac. de Ciencies, Torre C5-parell, 2a planta, 08193, Barcelona, Spain}
\altaffiltext{11}{ISOC, ESA/ESAC, Urb.~Villafranca del Castillo, P.O.~Box 50727, 28080 Madrid, Spain}
\altaffiltext{12}{Department of Physics, University of Alberta, Room 238 CEB, Edmonton, AB T6G 2G7, Canada}
\altaffiltext{13}{Steward Observatory, University of Arizona, Tucson, AZ 85721, U.S.A.}
\altaffiltext{14}{Astronomical Institute, Utrecht University, PO Box 80 000,3508 TA, Utrecht, The Netherlands}

\begin{abstract} \noindent We introduce the Galactic Bulge Survey
  (GBS) and we provide the \chan\, source list for the region that has
  been observed to date. Among the goals of the GBS are constraining
  the neutron star equation of state and the black hole mass
  distribution via the identification of eclipsing neutron star and
  black hole low--mass X--ray binaries. The latter goal will, in
  addition, be obtained by significantly enlarging the number of black
  hole systems for which a black hole mass can be derived. Further
  goals include constraining X--ray binary formation scenarios, in
  particular the common envelope phase and the occurrence of kicks,
  via source-type number counts and an investigation of the spatial
  distribution of X--ray binaries, respectively.  The GBS targets two
  strips of $6\degr\times1\degr$ (12 square degrees in total), one
  above ($1\degr<b<2\degr$) and one below ($-2\degr<b<-1\degr$) the
  Galactic plane in the direction of the Galactic Center at both
  X--ray and optical wavelengths. By avoiding the Galactic plane
  ($-1\degr<b<1\degr$) we limit the influence of extinction on the
  X--ray and optical emission but still sample relatively large number
  densities of sources. The survey is designed such that a large
  fraction of the X--ray sources can be identified from their optical
  spectra.  The X--ray survey, by design, covers a large area on the
  sky while the depth is shallow using 2~ks per \chan\, pointing.  In
  this way we maximize the predicted number ratio of (quiescent)
  low--mass X--ray binaries to Cataclysmic Variables. The survey is
  approximately homogeneous in depth to an 0.5-10 keV flux of
  7.7$\times 10^{-14}$ \flx. So far, we have covered about two--thirds
  (8.3 square degrees) of the projected survey area with \chan\,
  providing over 1200 unique X--ray sources.  We discuss the
  characteristics and the variability of the brightest of these
  sources.
\end{abstract}


\section{Introduction} 

\subsection{Multi-wavelength observations of X-ray sources}
X--ray observations are excellent probes of coronally active and
accreting sources. Whereas studies of X--ray sources in our
Galaxy have to a large extent focussed on bright systems,
investigations of fainter source classes have typically been done in
the Galactic Center (e.g.~\citealt{2003ApJ...589..225M}). There,
however, crowding and extinction make it more difficult to identify
the correct optical and/or infrared counterparts for large fractions
of the sources (e.g.~\citealt{2009arXiv0907.1935M}).

It is clear from previous surveys that multi--wavelength observations are vital
for classifying faint X--ray sources since X--ray spectral information
alone is rarely sufficient. Classification is important for various
science goals as we outline below. 

The Galactic Bulge Survey (GBS) we present in this paper was designed
to allow multi--wavelength observations of the detected X--ray
sources. The GBS consists of \chan\, and optical imaging of two strips
of $6^\circ \times 1^\circ$, one centered $1.5^\circ$ above the
Galactic plane and the other $1.5^\circ$ below the plane. We have
chosen this area as the source density is still high, but by excluding
$|b|<1^\circ$ we avoid the regions that are most heavily affected by
extinction and source confusion (see Figure~\ref{changbs}). In
Table~\ref{tab:reference} we provide a reference table for the acronyms
used in this paper.

\begin{table}
  \caption{Reference table for the acronyms used in this paper.}
\label{tab:reference}
\begin{center}
\begin{tabular}{ll}
ACIS & Advanced CCD Imaging Spectrometer \\
AGN & active galactic nucleus\\
AM CVn star & AM Canum Venaticorum star\\
ASCA & Advanced Satellite for Cosmology and Astrophysics\\
BB & black body\\
BH & black hole\\
Brems & Bremsstrahlung\\
BSC & bright source catalog\\
CALDB & calibration database\\
ChaMPlane & Chandra multi-wavelength plane survey\\
CTIO & Cerro Tololo Inter-American Observatory \\
CV & cataclysmic variable\\
EoS & equation of state\\
GBS & Galactic Bulge Survey\\
GLIMPSE & Galactic Legacy Infrared Mid-plane \\
 &Survey Extraordinaire\\
HMXB & high-mass X-ray binary\\
HR & hardness ratio\\
HRI & high resolution imager\\
ID & identification \\
IP & intermediate polar\\
LMXB & low-mass X-ray binary\\
NS & neutron star \\
PSPC & position sensitive proportional counter\\
RASS & ROSAT all sky survey\\
RS CVn star & RS Canum Venaticorum star\\
UCXB & ultra-compact X-ray binary\\
UKIDSS & UKIRT Infrared deep sky survey\\
UKIDSS/GPS & UKIDSS Galactic plane survey\\
VVV & Vista Variables in the Via Lactea\\
W UMa star & W Ursae Majoris star\\
\end{tabular}
\end{center}
\end{table}

\subsection{Goals of the GBS: compact object masses}
The discovery of a large sample of X--ray sources will allow us to
identify rare X--ray binaries such as quiescent eclipsing black hole
(BH) and neutron star (NS) low--mass X--ray binaries (LMXBs) and
ultra--compact X--ray binaries (UCXBs; LMXBs that have orbital periods
$<$1 hour). For dynamical mass measurements in LMXBs one needs to
measure three parameters: the radial velocity amplitude of the
companion star ($K_2$), the ratio between the mass of the companion
star and the NS or BH mass ($q$) and the inclination ($i$). A
measurement of the rotational broadening of the stellar absorption
lines ($v\sin i$) combined with $K_2$ gives this determination of $q$.
The reason is that in Roche lobe overflow systems, like LMXBs, the
companion star is tidally forced to co--rotate with the binary orbit
(\citealt{1988ApJ...324L..71T}).

The system inclination can be determined through modelling of
ellipsoidal variations caused by distortion of the companion star
using multi--color optical light curves
(e.g.~\citealt{2000A&A...364..265O}, \citealt{2010ApJ...710.1127C}).
In systems with favorable viewing angles, the (X--ray) eclipse
duration can be used to accurately determine the inclination
(\citealt{1985MNRAS.213..129H}). Since the inclination is constrained
by the geometry, mass measurements in eclipsing systems are
independent of the modelling that lies behind inclinations derived
from ellipsoidal variations.  Quiescent eclipsing systems are prime
targets for optical mass measurements. Such mass measurements provide
constraints on the NS equation of state (EoS) and on the dividing line
between NS and BH systems (e.g.~\citealt{2010ApJ...725.1918O}).
Constraining the NS EoS remains one of the ultimate goals for NS
studies (\citealt{2004Sci...304..536L} and references therein) and
mass measurements of BH systems would improve our estimates of the
stellar mass BH mass--distribution with implications for supernova and
Gamma--ray burst modelling. Furthermore, the BH sample selected in
quiescence, as in the GBS, is not susceptible to potential X--ray
outburst duty cycle based selection effects that occur when selecting
BHs after they have become X--ray bright and returned to quiescence,
which is now common practise. Finding eclipsing LMXBs with optical or
near-infrared counterparts bright enough that optical or
near--infrared spectroscopy can provide accurate mass measurements is
the principal goal of the GBS.

\subsection{Goals of the GBS: binary formation and evolution}
The second main goal of the GBS is to study the origin and evolution
of the population of X--ray binaries.  By comparing the observed
number of sources per source class (e.g.~CVs and LMXBs) with those
that have been predicted on the basis of population synthesis models,
one can constrain binary evolution models, in particular the nature of
the common envelope phase. Most compact binary sources responsible for
high energy phenomena went through one or two phases of
common--envelope evolution, such as CVs, AM Canum Venaticorum stars
(AM CVns) and UCXBs (\citealt{2000ARA&A..38..113T}). However, that
phase is not yet well understood. Compact binaries have much lower
orbital energy and angular momentum than the progenitor binary that
contained giants (\citealt{1976IAUS...73...75P}).  The binary
semi--major axis is thought to shrink mainly during a phase of
unstable mass transfer and ejection, i.e., the spiral--in. If the
outcome of this process is derived by assuming that the change in
orbital energy is enough to eject the giant's mantle, the predicted
properties do not match the observations of double white dwarf
binaries. These properties can be matched with the assumption that the
giant's mantle is ejected, carrying a fixed fraction of the total
angular momentum (\citealt{2000A&A...360.1011N}), but this begs the
question how the required energy is provided. A more complete
theoretical description is required that takes into account both
energy and angular momentum.

From the properties of single radio pulsars it has been inferred that
they receive a kick at birth, due to asymmetries in the supernova
explosion (\citealt{1994Natur.369..127L}). This has important
consequences for the evolution and properties of X--ray binaries
(e.g.~\citealt{1995MNRAS.274..461B}, \citealt{1996ApJ...471..352K}) so in
principle X--ray binaries can be used to constrain the properties of
the kick.  In particular, the question is whether black holes also
receive a kick at formation and, if so, what its properties are
(e.g.~\citealt{1996ApJ...473L..25W}, \citealt{2004MNRAS.354..355J},
\citealt{2005ApJ...625..324W}, \citealt{2009MNRAS.394.1440M}).

Four main steps must be followed in binary population synthesis
modeling (\citealt{lrr-2006-6}).  First, a set of initial conditions
must be chosen.  Key parameters which must be set include the initial
mass function, the initial distribution of mass ratios in binaries,
the initial binary fraction, and the initial distribution of orbital
periods.  Secondly, any evolution with time of these initial
conditions such as due to changes in star formation rate should be
taken into account.  Next, a recipe for (binary) stellar evolution must be
specified.  The major uncertainties in this recipe include the
aforementioned uncertainties in common envelope evolution;
uncertainties in the neutron star and black hole natal kick
distributions; and uncertainties in mass and angular momentum loss.
Finally, a recipe must be developed for converting accretion rates,
typically calculated on timescales of centuries or longer, into
instantaneous observables such as the X--ray luminosity.  Key
uncertainties in this step largely involve understanding the spectra
of accreting sources at different accretion rates, and understanding
how disk instabilities will affect the distribution of luminosities of
those sources which undergo disk accretion.

To make progress on these issues we envisage a two--pronged approach
that includes detailed studies of individual systems on the one hand,
and of the population on the other hand. Any viable evolutionary
scheme must be able to reproduce the specific properties (such as
component masses, orbital period, age, system velocity) of each
observed individual system. Any viable evolution scheme must also
reproduce the distributions of and correlations between these
properties in the population of X--ray binaries.  Observationally this
requires the discovery of large homogeneous samples of CVs and
(ultra--compact) X--ray binaries, and the detailed follow--up of a
number of individual systems.

The large number of sources will also allow us to investigate the
spatial distribution of LMXBs and this provides input to the LMXB
formation scenarios. \citet{2004MNRAS.354..355J} found that there is a
significant excess of LMXBs at $-10^\circ<l<0^\circ$ with respect to
$0^\circ<l<10^\circ$.  \citet{2008Natur.451..159W} found an
asymmetric distribution of the 511 keV line emission using INTEGRAL
data which they suggested was due to an asymmetric LMXB distribution
(although see \citealt{2009MNRAS.392.1115B}). However, any spatial
non--uniformity in the Galactic birth distribution of LMXBs is thought
to be washed out by the natal kick imparted on the NS in the LMXB
formation models presented by \citet{1983adsx.confQ.303V} and
\citet{1998ApJ...493..368K}. On the other hand, the LMXB evolutionary
model, involving triple star evolution, discussed by
\citet{1986MNRAS.220P..13E} provides a channel for the formation of
LMXBs without a large kick velocity. In general, except for formation
scenarios involving either direct collapse to a BH or an accretion
induced collapse, a Blaauw kick should be imparted regardless of the
type of compact object formed during the supernova event
(\citealt{1961BAN....15..265B}).  Evidence for velocity kicks is found
in the $z$--distribution of LMXBs (\citealt{vawh1995};
\citealt{2004MNRAS.354..355J}) and in the velocity distribution of
radio pulsars (\citealt{1994Natur.369..127L};
\citealt{1997MNRAS.291..569H}). An improved spatial distribution of
LMXBs in the Bulge will help to test if such kicks happen commonly or
rarely.

\subsection{Previous X--ray surveys of the bulge}
Previous X--ray surveys of the Galactic bulge provide some general
expectations for our survey.  The Advanced Satellite for Cosmology and
Astrophysics (ASCA) observed the central region of our Galaxy with $|
l |<45^\circ$ and $| b | <0.4^\circ$ in the 0.7--10 keV band, albeit
with a resolution of 3\arcmin\, which led to a large number of
unclassified sources (\citealt{2001ApJS..134...77S}). The ASCA survey
discovered 163 sources down to a flux of $\approx 3\times
10^{-12}$~\flx\, which as a group have properties in common with
Cataclysmic Variables (CVs), high--mass X--ray binaries (HMXBs),
quiescent LMXBs and Crab-like pulsars. A recent paper by
\citet{2010arXiv1011.3295A} finds a number of massive stars within the
ASCA Galactic plane sources. The XMM--{\it Newton} Galactic plane
survey covered 3 square degrees along the Galactic plane down to $F_X$
(2-10 keV) $2\times10^{-14}$~\flx, finding roughly one--third to be
soft sources suggestive of nearby coronally active stars
(\citealt{2004MNRAS.351...31H}), while two--thirds were hard, absorbed
sources. The latter group is dominated by active galactic nuclei (AGN)
but it also has a substantial Galactic component, especially between
$10^{-12}>F_X>10^{-13}$~\flx.  Optical identifications of 30 sources
with $F_X>7\times10^{-14}$~\flx\, led to the classification of 16-18
coronally active stars, 3 CVs, and 2 LMXB candidates
(\citealt{2010A&A...523A..92M}).  Finally, spectroscopic follow up of
medium--duration \chan\, observations of numerous Galactic plane
fields (the ChaMPlane survey, \citealt{2005ApJ...635..920G};
\citealt{2006ApJS..163..160R}) finds roughly one--third to be soft
sources, largely stars, and two--third to be hard sources, including
many AGN and a few CVs.

In this paper we describe the modelling that provides rough estimates for the
number of sources that we expect in the GBS ($\S 2$). The \chan\, X--ray data,
the X--ray source list, and analysis of the X--ray source properties are
described in $\S 3$. A brief outline of the multi--wavelength imaging data is
given in $\S 4$, followed by a comparison of ROSAT sources in the GBS area with
the \chan\,--discovered sources is $\S 5$. We end with a short summary ($\S 6$)
and outlook ($\S 7$).

\begin{figure} 
\includegraphics[angle=0,width=8cm,clip]{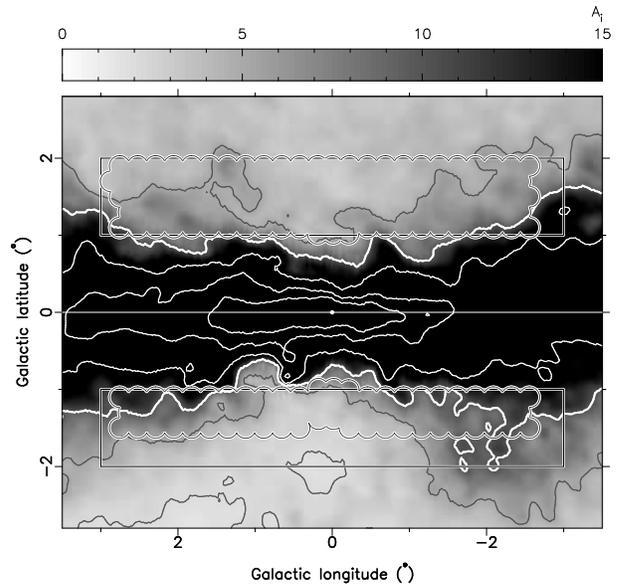}
\caption{The large black plus white rimmed boxes indicate the two
  $6^\circ\times1^\circ$ GBS fields in Galactic coordinates. The thick
  white contour made-up of small circles each of a
  14\arcmin\,diameter, indicate the area covered by our \chan\,
  observations until 2010. The grey scale image and contours depict
  the total absorption in the Sloan $i^\prime$--band filter,
  $A_{i^\prime}$, estimated from the \textsc{Cobe} dust maps
  \citep{1998ApJ...500..525S}. The contours are at $A_{i^\prime}$
  values of 2, 5 (both in dark grey), 10, 20, 50 and 100 (in white).}
\label{changbs} \end{figure}

\section{Source number estimates}

We used a simple method to estimate roughly the number of X--ray
sources in the Bulge area and based on this we devised an optimal
observing strategy to best meet our goals. For each source class we
used estimates of their typical optical ($i^\prime$-band) and X--ray
brightness as well as local space densities or total number of objects
in the Galaxy without following their formation or evolution.  We
distribute these sources in the Galaxy according to the model for the
star formation history from \citet{2004MNRAS.349..181N}, which is
based on the Milky Way disk formation simulation of
\citet{1999MNRAS.307..857B} with an added Bulge component. We do not
take the influence of kicks for neutron star systems into account. For
each object we determine the interstellar absorption from the
\citet{1998ApJ...500..525S} dust maps, assuming the dust is evenly
distributed (i.e.~homogeneous) between the Earth and the position
where the line of sight leaves the dust layer for an assumed dust
height above and below the Galactic plane of 120 pc. For the optical
we use $A_V = 3.1 E(B-V)$ and $A_{i^\prime} = 0.6 A_V$. We further use
$A_K = 0.1 A_V$. For the X--ray absorption we distinguish hard sources
(with an assumed power law spectrum with photon index 2), soft sources
(where the assumed spectrum is a black body with temperature $0.25$
keV), and sources with a Bremsstrahlung spectrum with a temperature
of 2 keV. We link the absorption to the reddening via $N_{\rm H} =
0.179 \times 10^{22} A_V$ \citep{1995A&A...293..889P}.

\begin{figure} 
\includegraphics[angle=-90,width=8cm,clip]{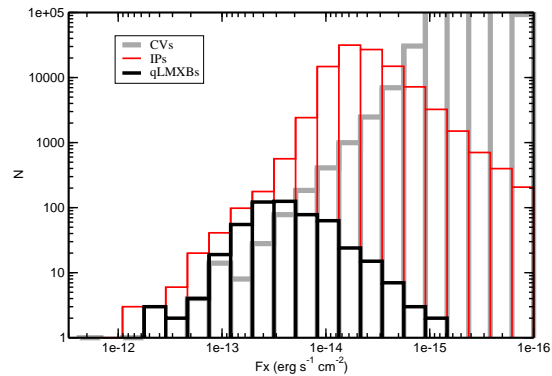}
\caption{The predicted number of non-magnetic CVs (thick blue lines), intermediate 
polars (thin red lines) and quiescent LMXBs (medium thickness black lines) 
in the GBS area as a function of source X--ray flux. The approximate flux 
limit of the GBS is $(1-3)\times 10^{-14}$~\flx, with the variations 
largely due to differences in source spectra. We expect that deeper 
\chan\, observations would increase the number of CVs much more strongly 
than they would increase the number of quiescent LMXBs.
 }
\label{cv-qlmxb} \end{figure}

By comparing the expected number of quiescent LMXBs with the number of
CVs as a function of X--ray flux (see Figure~\ref{cv-qlmxb}) we
decided on rather shallow \chan\, observations ($\approx 2$ ks per
pointing) yielding a limiting flux of $(1-3) \times
10^{-14}$~\flx. This limit is dependent on the source spectrum
including extinction. For the source number estimates we take a
general flux limit of 3$\times 10^{-14}$~\flx. Deeper observations
would mostly increase the number of detected CVs. This would make it
more difficult to classify the LMXBs among the sample of X--ray
sources. As mentioned above, we selected two strips of $6^\circ \times
1^\circ$, one centered at $1.5^\circ$ above and one centered at
$1.5^\circ$ below the Galactic Center (see Figure~\ref{changbs}). The
size and location are chosen in order to detect enough sources in each
class to achieve our science goals. The optical extinction is still
substantial in this region but relatively low compared to the Galactic
Center and Galactic plane. This will allow us to obtain optical
spectroscopic follow--up to a large fraction of the X--ray sources,
which is crucial for achieving our science goals. Going to a higher
latitude would facilitate identification as the crowding would be less
severe, but the source density drops off quickly, which would require
a larger surface area to detect a sufficient number of sources. The
chosen strips are thus a compromise between survey area size and
location.

\begin{table*}
  \caption{Estimated numbers of the various source categories. The columns
    describe: the source type (I), the assumed mean X--ray luminosity
    in \lum\, (II), the
    X--ray color (III), the assumed mean $i^\prime$--band
    absolute magnitude (IV), assumed $i^\prime-K$ intrinsic color (V), assumed
    total number of sources in the Galaxy
    (integer values) or the space densities ($\rho$ pc$^{-3}$) (VI), number of sources that
    should be detectable 
    both in the GBS X--ray as well as the optical survey (VII), number of
    sources detected in the GBS X--ray as well as in a $K$--band
    imaging survey of limiting magnitude of $K=18$ (VIII), number of
    sources detected in X--rays in the GBS area (IX), number of sources
    falling in the GBS area (X). Quiescent sources are denoted with
    'q'. 'BB' stands for a 0.25 keV 'black body'. 'Brems' stands for Bremsstrahlung.}
\label{tab:model}
\begin{center}
\begin{tabular}{lllrrrrrrr}
\hline
(I)   & (II) & (III) & (IV) &(V) &(VI)&(VII) &(VIII) & (IX) &(X)\\
\hline
LMXB & $10^{35}$  & Hard & 0 & 0 & 140 & 6 &  7 & 7 &  7 \\ 
qLMXB & $10^{33}$  & BB  & 5 & 2  & 10000 & 120 & 86 & 221 & 532 \\ 
UCXB & $10^{34}$  & Hard & 4 & 0 & 1000 & 32 & 3 & 56 &  58 \\ 
qUCXB & $10^{32}$  & Hard & 10 & 0 & 10000 & 1 & 0 &   8 &  605 \\ 
CV (non mag.) & $10^{31}$  & Brems & 7.5 & 0 &  $2 \times 10^{-5}$ &
62 & 61 &  62 & $1.4 \times 10^6$ \\
CV (IP) & $10^{32}$  & Brems & 8.5 & 0 &  $1.5 \times 10^{-6}$ & 152
& 5 &  525 & $7.7 \times 10^4 $\\ 
RS CVn & $10^{31}$  & Hard & 2.5 & 1 & $1 \times 10^{-4}$ & 596 &  596
&  596 & $1.3 \times 10^6$ \\
W UMa & $5\times 10^{30}$ & Hard & 4.5 & 2 & $7.5\times 10^{-5}$ & 160
& 160 & 160 & 2.3$\times10^6$ \\ 
Be X-ray binaries  & $10^{34}$& Hard & 0 & 0 & 500 & 9 & 9 & 10 & 10 \\  
\hline
Total &  &  &  &   &  & 1142 &  & 1648 &  \\  

\end{tabular}
\end{center}
\end{table*}

The number of sources expected in the GBS area and the assumptions
made in the population estimates are given in Table~\ref{tab:model}.
The values for the assumed X--ray luminosity, $i^\prime$--band
absolute magnitude and optical extinction A$_V$ in
Table~\ref{tab:model} are averages. In the modelling we assumed a
distribution of X--ray luminosities using a Gaussian distribution, with
$\sigma=0.5$ and the average normalized to
1. For the $i^\prime$--band absolute magnitude we apply a Gaussian
smoothing with $\sigma$=1 magnitude. Finally, we use a Gaussian
smoothing in the assumed A$_V$ with $\sigma=0.2$ magnitude. 

We expect to discover Roche lobe overflow NS and BH X--ray binaries, both with
main sequence donors (LMXB) as well as UCXBs.  Most LMXBs will be in
quiescence.  The numbers for the LMXBs are based on a very simple estimate of
$\approx$140 persistent systems with $L_\mathrm{X}>10^{35}$\,erg\,s$^{-1}$ in
the Galaxy (\citealt{2002A&A...391..923G}), a number ratio of quiescent to
persistently bright/active LMXBs of $\sim70$, and the prediction that the number
of UCXBs may be equal to the number of non-degenerate LMXBs
(\citealt{2004ApJ...603..690B}). To estimate the number ratio of quiescent to
active LMXBs we use results of surveys of Galactic globular clusters as input. 
There, roughly 10 times as many quiescent LMXBs are identified (by their soft,
blackbody-like spectra) as active LMXBs (\citealt{2003ApJ...598..501H}).  Since
fainter quiescent LMXBs tend to be dominated by their hard, power-law component
rather than the blackbody component (e.g.~\citealt{2004MNRAS.349...94J} and
\citealt{2008AIPC..983..519J}), it has been suggested that half of all quiescent
LMXBs are missed in these globular cluster surveys
(\citealt{2005ApJ...622..556H}).  However, it is uncertain whether globular
cluster LMXBs are similar in their duty cycles to Galactic LMXBs, as they are
formed in different ways.

An alternative way to estimate the number ratio between active and
quiescent LMXBs comes from the total number of LMXBs and the estimated
X--ray binary lifetime.  \citet{1997A&A...321..207P} compute that
there are $10^{-5}$ NS binaries formed per year for most reasonable
values of the common envelope parameter.  Assuming a typical lifetime
as an X--ray binary of approximately 1 Gyr, then there are $\approx
10^4$ systems in the Galaxy.  More recently,
\citet{2006MNRAS.369.1152K} find a similar number from their
evolutionary population synthesis calculations.  Taking that there are
$\approx$140 active X--ray binaries at any given time, one in seventy
are active at any given time. We take this 70 as our assumed ratio
between active and quiescent LMXBs in our number estimates.

In an X--ray selected sample such as this, another major population of
sources that we will pick-up is CVs and of these we will in particular
detect magnetic Intermediate Polar systems (IPs). Their number
estimate is based on the observed IP and non-magnetic CV space
densities of $\sim1.5\times10^{-6}$ and $2\times10^{-5}$\,pc$^{-3}$
(\citealt{1984ApJS...54..443P}; \citealt{1990ApJ...364..251H},
\citealt{2007MNRAS.382.1279P}, \citealt{2008ApJ...675..373R}). The
discovery of a large number of IPs with INTEGRAL
(\citealt{2010MNRAS.401.2207S}) may be indicating that the fraction of
IPs among the CVs is substantially larger than was previously
thought. In addition, there will be a large number of active (binary)
stars such as RS CVn stars. The expected number of background AGN is
25 (see \citealt{2005ApJ...635..214E}).  We might also find a few
ultra-compact double white dwarf systems (AM CVns).

The expected number of W UMa--like sources is calculated using the
space density of 7.5$\times 10^{-5}$ pc$^{-3}$ with ${\rm
  M_{i^\prime}\approx 4.5}$ and an average X--ray luminosity of
$5\times10^{30}$\lum\, (\citealt{1998AJ....116.2998R}). Finally, the
expected number of Be -- X--ray binaries is estimated using an
estimated total number of systems of 500 in our Galaxy, with an ${\rm
  M_{i^\prime}}=0$ and a hard power law spectrum with an average
luminosity of ${\rm L_X=1\times 10^{34}}$\lum. We expect 10 in the GBS
area out of which 9 will be detectable in the $i^\prime$--band as well
as the $K$--band.

These numbers are uncertain and calibrating the number densities of
sources using the GBS is one of the principal objectives. We will
compare our identifications with the predicted numbers of binaries in
each category and thus place strong constraints on important
uncertainties in the theory of binary evolution, such as the
common--envelope phase and the existence and magnitude of kicks
imparted on NSs and BHs.

Of the predicted LMXBs we assume that 10 per cent will have BH
accretors. This number is highly uncertain but 10 per cent is roughly
the number ratio between NS and BH systems observed in HMXBs (see
\citealt{2003MNRAS.339..793G} and references therein). Some authors
have claimed that the ratio between NS and BH systems should be closer
to 1 (\citealt{1992ApJ...399..621R}), whereas others indeed find
values closer to 10 per cent (\citealt{1997A&A...321..207P};
\citealt{2006MNRAS.369.1152K}).  Assuming 10 per cent would lead to
$\sim 25$ BH LMXBs in the GBS which will approximately double the
population of known Galactic BHs for which a dynamical mass
measurement is possible (currently $\sim$20,
\citealt{2006csxs.book..157M}). Eclipsing BH LMXB systems should
exist, but have not yet been found. It has been proposed that they are
too weak to be detected by current X--ray all sky monitors because
they are obscured behind the accretion disk rim during outburst
(\citealt{2005ApJ...623.1017N}). If so, they should turn up in our GBS
in quiescence, when the disk is expected to be much thinner. The
number of eclipsing sources depends on the distribution of the mass
ratio between the accretor and the donor star (see
e.g.~\citealt{1985MNRAS.213..129H}). For mass ratios $q \approx 0.3$
approximately 20-25 per cent of the 200 new quiescent LMXBs we expect
to discover should be eclipsing of which a handful could be BHs.

\begin{figure*} 
\includegraphics[angle=-90,width=16cm,clip]{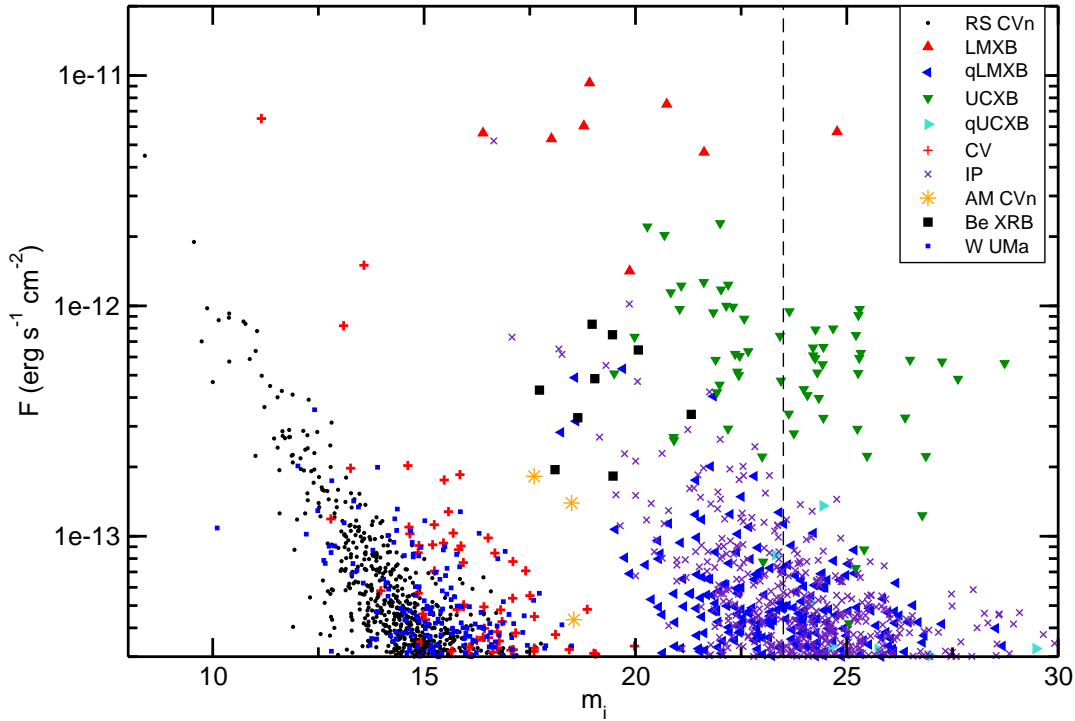}
\caption{Predicted source X--ray flux, F$_X$, as a function of $i^\prime$--band magnitude for the
 sources predicted to be most numerous in our GBS survey. The vertical
dashed line indicates the approximate $i^\prime$--band magnitude limit
of our optical imaging.}
\label{gbs-i-fx} \end{figure*}

Overall, this modelling can be summarized in a figure with X--ray flux
as a function of optical $i^\prime$--band magnitude for the main
components of the expected sources (Figure~\ref{gbs-i-fx}).  The large
majority of bright optical sources with $i^\prime\approxlt 16$
associated with an X--ray source in the GBS area is expected to be an
RS CVn. Similarly, the majority of X--ray bright sources
(F$_X>1\times10^{-12}$~\flx\, with a optical counterpart with
$18<i^\prime<22$ is expected to be an active LMXB or UCXB. Finally,
most quiescent LMXBs and UCXBs will have faint optical counterparts
($i^\prime>21$).

Clearly, other source types are expected, for instance single X--ray active nearby
G, K or M stars (cf.~\citealt{2004A&A...417..651S}) and Algol 
sources.  However, the expected number of sources is small for the
Algol and nearby single stars, so we do not provide detailed estimates
for these classes. 

In order to roughly assess the accuracy of the aforementioned modelling, we
compare the number of sources detected in the single, 1 Ms deep \chan\,
observation of a small part of 20.6 square arcminutes of the GBS area presented
by \citet{2009Natur.458.1142R} with the number of sources we predict if one were
to observe the whole GBS area to that depth of $1\times 10^{-16}$ \flx. We
expect to find 3 million sources based on our source number estimates, whereas
scaling the observed 20.6 squared arcminutes to 12 square degrees we would
expect to find about 1 million sources. The main difference between these two
numbers probably stems from uncertainties in scaling from the 20.6 square
arcminute area to a 12 square degree field such as those caused by the
difference in extinction between the GBS area with respect to that in the area
studied by \citet{2009Natur.458.1142R}, as well as of course uncertainties in
our modelling. All in all, the agreement within a factor of 3 between the number
of sources we predict and what was found is not unsatisfactory.

\section{X--ray observations} 

We have obtained observations with the \chan\, X--ray observatory
(\citealt{2002PASP..114....1W})  covering, to date, about
two--thirds of the total area of twelve square degrees that we
envisage for the GBS.

In Figure~\ref{changbs}, the wiggles indicate the composite outline of
each circular field of view of 14\arcmin\, diameter of the individual
\chan\, observations obtained to date covering the GBS area. The
\chan\, observations have been performed using the I0 to I3 CCDs of
the Advanced CCD Imaging Spectrometer (ACIS) detector
(\citealt{1997AAS...190.3404G}; ACIS--I). The observation
identification (ID) numbers for the data presented here are \dataset
[ADS/Sa.CXO#DefSet/GBS] {8643--8774
and 9977--10024}. We reprocessed and analyzed the data using the {\sc
  CIAO 4.2} software developed by the \chan\, X--ray Center and
employing {\sc CALDB} version 4.3. The data telemetry mode was set to
{\it very faint} for all observations except that with ID 8687 since
there was a bright source present in archival ROSAT observations. For
that observation we used the standard {\it faint} mode.

We also reprocessed and reanalyzed the data that was obtained as
part of the Bulge Latitude Survey (\citealt{2005ApJ...635..920G};
\citealt{2009ApJ...706..223H}; observation IDs 7160-7162, 7166-7168,
8199-8204, 9562-9564) that falls in the area we target in the GBS
using exactly the same reduction and analysis as for our GBS
observations. Because the original Bulge latitude survey
\chan\, observations have an exposure time of approximately 15~ks, we
selected 2 ks segments to allow for a comparison with our GBS
observations. We selected data stretches of 2 ks length taking
the start time of the observation from the header of the data, plus
100 seconds as the starting point of the 2 ks stretches. These data
were also obtained using the {\it very faint} mode.

The {\it very faint} mode provides 5$\times$5 pixel information per
X--ray event. This allows for a better screening of events caused by
cosmic rays.  In our analysis we selected events only if their
energy falls in the 0.3--8 keV range.

We used {\sc wavdetect} to search for X--ray sources in each of the
observations using data covering the full 0.3--8, the 0.3-2.5 and the
2.5-8 keV energy band separately. We set the {\sc sigthresh} in {\sc
  wavdetect} to 1$\times 10^{-7}$, which implies that for a background
count rate constant over the ACIS-I CCDs there would be 0.1 spurious
source detection per observation as about $1\times 10^6$ pixels are
searched per observation. However, as we explain below, we applied
additional selection criteria.  This lowers the number of spurious
sources.

The resulting {\sc SIGNI} column in the output list of detected
sources provided by {\sc wavdetect} is an estimate of the photon flux
significance, not of the detection significance. Instead, we retained
all sources for which Poisson statistics indicates that the
probability of obtaining the number of detected source counts by
chance given the expectation for the local background count rate is
lower than 1$\times 10^{-6}$. This would be equivalent to a $>5\sigma$
source detection in Gaussian statistics. Next, we deleted all sources
for which {\sc wavdetect} was not able to provide an estimate of the
uncertainty on the right ascension [$\alpha$] or on declination
[$\delta$] as this indicates often that all counts fell in 1 pixel
which could well be due to faint afterglows events caused by cosmic
ray hits. In addition, we impose a 3 count minimum for source
detection as \citet{2005ApJS..161....1M} simulated that in their
XBootes survey with 5 ks ACIS--I exposures, 14 per cent of the 2 count
sources were spurious (note that this percentage will probably be
lower for our GBS exposures of 2 ks).

Since our \chan\, observations were designed to overlap near the
edges, we searched for multiple detections of the same source either
in one of the energy sub-bands or in the full energy band. We consider
sources with positions falling within 3\arcsec\, of each other likely
multiple detections of the same source. Finally, we inspected the
source list and found two spurious sources caused by the bright
read-out trail and piled-up core of source \# 1 (see
Table~\ref{srclist}).

In total we detected 1234 distinct sources in the area indicated with
circles in Figure~\ref{changbs}, including sources detected in the
Bulge Latitude Survey area that fall within our GBS area. The source
list is given in Table~\ref{srclist} and provides information on
$\alpha$, $\delta$, the error on $\alpha$ and $\delta$, total number
of counts detected, the observation ID of the observation resulting in
the detection and the off-axis angle at which the source is detected.
The error on $\alpha$ and $\delta$ are the error provided by {\sc
  wavdetect}, it does not take  into account the typical \chan\, bore--sight
uncertainty of 0.6\arcsec\, (90 per cent confidence).

We provide individual \chan\, source names, however, for briefness we
use the source number in Table~\ref{srclist} to indicate which source
we discuss in this paper. For the error $\sigma_N$ on the detected
number of counts $N$, \citet{2005ApJS..161..271G} give $\sigma_N =
1+\sqrt{N+0.75}$ after \citet{1986ApJ...303..336G}. To allow for an
rough, easy calculation of the source flux based on the detected
number of source counts we give the conversion factor for a source
spectrum of a power law with photon index of 2 absorbed by $N_{\rm
  H}=1\times 10^{22}$ cm$^{-2}$: $7.76\times
10^{-15}$~\flx\,photon$^{-1}$.

\renewcommand{\arraystretch}{2.0}
\begin{center}
\begin{longtable*}{ccccccccccc}
\caption{PLACEHOLDER, FIRST TEN ENTRIES ONLY! See http://www.sron.nl/$\sim$peterj/gbs or the electronic version in ApJS
for the full source list. The GBS X--ray source list providing the GBS source name, the source number as used in this
paper,  $\alpha$, $\delta$, the error on $\alpha$ and $\delta$, total number of counts detected, the observation ID
of the observation resulting in the detection, the off-axis angle at which the source is detected and the hardness
ratio (HR) for sources detected with more than 20 counts. The HR is calculated for the detection where the off-axis 
angle was smallest if the source was detected multiple times.}
\label{srclist}\\
\hline
Source & \# & $\alpha$& $\delta$ & Error $\alpha$ &
Error $\delta$ & \#  & Obs & Off-axis & \# of & HR \\[0.1mm]
name &  & (degrees) & (degrees) & (arcsec) & (arcsec) & (cnt) & ID &
angle (\arcmin) & detec. & \\[0.1mm]

CXOGBS J175024.4-290216&1 & 267.60182588  & -29.037885415 & 0.034 & 0.011 & 3391 &  8709 &  8.99 & 1   &\nodata 		   \\[0.1mm]
CXOGBS J173728.3-290802&2 & 264.36831021  & -29.133892744 & 0.058 & 0.084 & 2191 &  8691 &  7.63 & 2   &\nodata 		   \\[0.1mm]
CXOGBS J174042.8-281808&3 & 265.17839123  & -28.302224391 & 0.048 & 0.033 & 1850 &  8687 &  5.54 & 4   &\nodata 		   \\[0.1mm]
CXOGBS J173931.2-290952&4 & 264.88008702  & -29.164675805 & 0.104 & 0.072 &  238 &  8679 &  5.63 & 1   &-0.78 $\pm$0.09\\[0.1mm]
CXOGBS J174009.1-284725&5 & 265.03805780  & -28.790455138 & 0.055 & 0.054 &  157 &  8677 &  3.41 & 1   & 0.50 $\pm$0.10\\[0.1mm]
CXOGBS J174445.7-271344&6 & 266.19074704  & -27.229022849 & 0.065 & 0.047 &  153 &  8647 &  3.20 & 1   &-0.25 $\pm$0.09\\[0.1mm]
CXOGBS J173826.1-290149&7 & 264.60910401  & -29.030389506 & 0.160 & 0.155 &  150 &  8690 &  6.65 & 2   &-0.89 $\pm$0.12 \\[0.1mm]
CXOGBS J173508.2-292957&8 & 263.78448781  & -29.499426909 & 0.119 & 0.073 &  138 &  9997 &  4.78 & 1   &-0.39 $\pm$0.10 \\[0.1mm]
CXOGBS J173508.3-292328&9 & 263.78498489  & -29.391224268 & 0.141 & 0.135 &  134 &  9996 &  5.95 & 1   &-0.64 $\pm$0.12 \\[0.1mm]
CXOGBS J173629.0-291028&10 & 264.12099622  & -29.174663351 & 0.322 & 0.180 &  122 &  9995 &  8.34 & 3  &-0.76 $\pm$0.16 \\[0.1mm]

\end{longtable*}
\end{center}
\renewcommand{\arraystretch}{1.0}

\subsection{Sources detected multiple times}

In the trade--off between a homogeneous survey depth and total survey
exposure time we used 7\arcmin\, as an effective radius of the \chan\,
field of view in designing the survey. The size and shape of the point
spread function for off--axis angles larger than 7\arcmin\, has
degraded such that many optical and/or near--infrared stars will fall
inside the X--ray error circle, even for sources detected with more
than 10 counts, making multi--wavelength follow--up more difficult
(cf.~\citealt{2005ApJ...635..907H}). As noted above, this yields the
possibility of multiple detection of sources discovered in the overlap
regions. Indeed, 105 sources are detected more than once, where we
have taken source positions within 3\arcsec \, of each other as
multiple detections of the same source. Out of these 105 sources, 95
sources are detected two times, 9 sources are detected three times and
1 source is detected 4 times. The properties that we list in
Table~\ref{srclist} for these sources are those of the detection that
gave rise to the largest number of X--ray counts. In
Table~\ref{srclist} we also list the number of times that sources are
detected. In addition, we found that although \chan\, source \# 33 and
\# 230 are 4.8\arcsec\, apart, they are conceivably two detections of
the same source. Both detections are rather far off-axis (5.2\arcmin\,
and 8.8\arcmin, respectively) explaining the relatively large
uncertainty in both measurements of the position.

\subsection{Detection probability vs off-axis angle}
 
\begin{figure} \includegraphics[angle=0,width=8cm,clip]{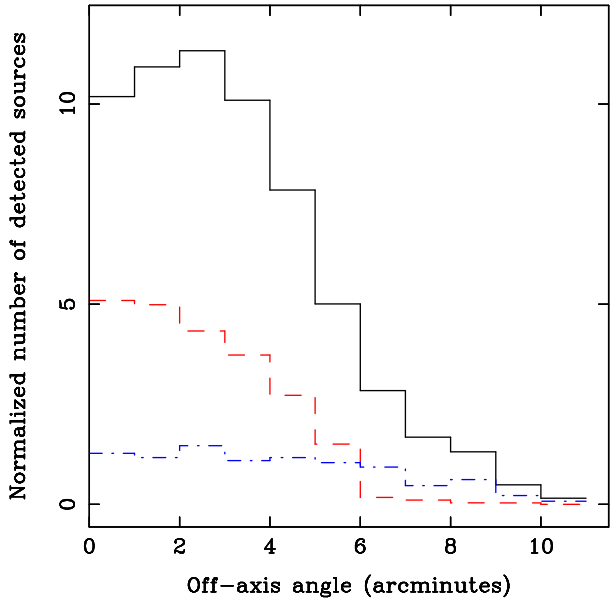}
  \caption{The number of detected sources normalized to area on the
    sky as a function of off-axis angle. The black solid line
    indicates all sources, the red dashed line indicates sources
    detected with only 3 X--ray counts and the blue dot-dash line are
    sources detected with 10 counts or more. From the black and red
    lines it is clear that the survey sensitivity to low count sources
    is higher when nearer to on-axis, as expected. The fact that the
    blue line is approximately flat up to 7\arcmin\, indicates that
    the GBS survey sensitivity is approximate constant for sources of
    10 counts or more. For a power law spectrum with photon index of 2
    and an N${\rm _H}=1\times 10^{22}$ cm$^{-2}$ this corresponds to a
    flux of 7.7$\times 10^{-14}$~\flx.}
\label{numsrcoffax} \end{figure}

\begin{figure} \includegraphics[angle=0,width=8cm,clip]{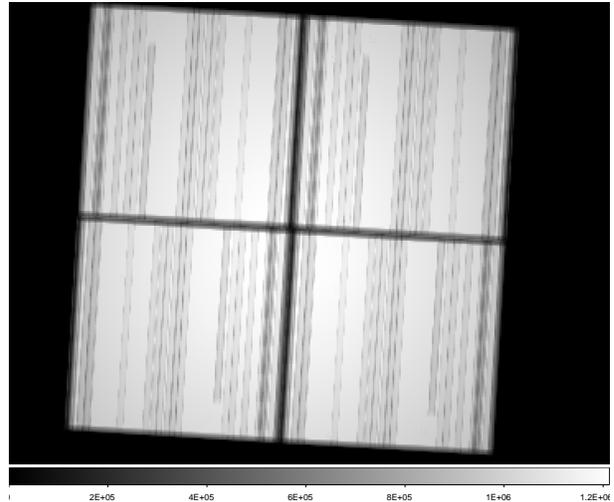}
  \caption{The exposure map for the observation with Obs ID 9999 for
    an assumed mono-chromatic source emitting 1 keV photons, showing
    the lower exposure for the gaps between the ACIS I CCDs. Lighter
    areas have a higher exposure on the sky (in terms of cm$^2$
    s$^{-1}$). Due to the satellite dithering some exposure is
    achieved even for the gaps. The vertical strips are caused by the
    presence of columns of pixels of reduced efficiency. The grey area
  is approximately 8\arcmin$\times$8\arcmin\, in size.}
\label{exposuremap} \end{figure}

We investigate the relationship between the number of sources detected
as a function of off-axis angle and the number of source counts. In
this way we can determine the approximate number of source counts
where the survey sensitivity for off--axis angles less than 7\arcmin\,
is constant. We find (Figure~\ref{numsrcoffax}) that when normalized
to the surface area, the number of detected sources drops as a
function of off-axis angle for angles larger than $\approx$2\arcmin.
The probable reason for the slightly lower number of sources detected
per unit surface area inside the 2\arcmin\, radius is that the chip
gaps between the 4 ACIS--I CCDs subtend a larger fraction of the total
solid angle within 2\arcmin\, than they do in the outer regions. That
area has a lower effective exposure time as can be seen in
Figure~\ref{exposuremap}. For source counts larger than 10 the
normalized number of detected sources is approximate constant for
off--axis angles less than 7\arcmin~(blue dot-dash line in
Figure~\ref{numsrcoffax}). Thus the GBS coverage is approximately
homogeneous down to source fluxes of $\approx7.7\times 10^{-14}$~\flx,
whereas the sensitivity is decreasing with off--axis angle for fainter
sources.

\subsection{X--ray spectral information}

\begin{figure} \includegraphics[angle=0,width=8cm,clip]{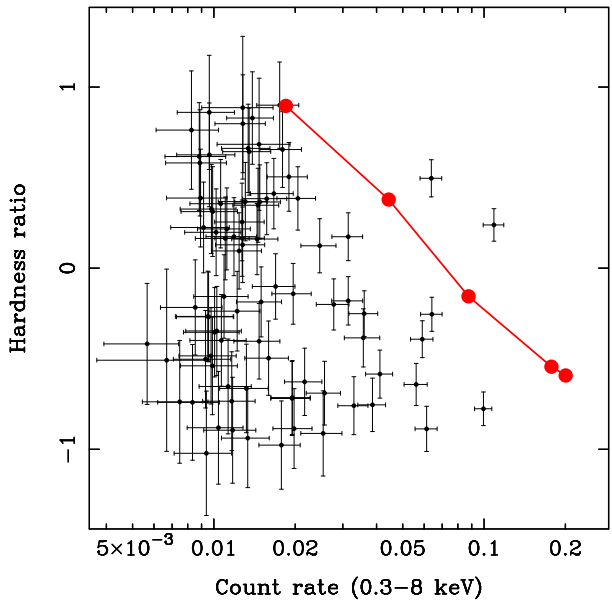}
  \caption{The hardness -- intensity diagram for the 86 sources for
    which more than 20 counts were detected in the GBS survey (we
    exclude the three brightest sources from the plot since their
    hardness ratio is strongly affected by effects of photon
    pile-up). To mitigate effects of small differences in exposure
    times we used count rates as a measure of intensity. The hardness
    is defined as the ratio between the count rate in the 2.5--8 keV
    minus that in the 0.3--2.5 keV band to the count rate in the full
    0.3--8 keV energy band. Hard sources fall in the top half and soft
    sources in the bottom half of this figure. The red line with
    large bullets shows the influence of the extinction N${\rm_H}$ on
    a power law spectrum with index 2 for a source count rate of 0.05
    counts s$^{-1}$ and N${\rm_H}$ of 0.01 $\times 10^{22}$
    cm$^{-2}$. The N${\rm_H}$ is increasing from bottom right to top
    left from (0.01, 0.1, 1, 3, 10)$\times 10^{22}$ cm$^{-2}$. }
\label{hardnessinten} \end{figure}

We extract source counts using circular source extraction regions of
10\arcsec.  Background extraction regions are annulli with inner and
outer radii of 15\arcsec\, and 30\arcsec, respectively. We plot the 89
sources for which we detected more than 20 counts in a hardness --
intensity diagram (Figure~\ref{hardnessinten}). To mitigate the
effects that small differences in exposure time across our survey can
have, we use count rates as a measure of intensity. We define the
hardness ratio as the ratio between the count rate in the 2.5--8 keV
minus that in the 0.3--2.5 keV band to the count rate in the full
0.3--8 keV energy band (after~\citealt{2004ApJS..150...19K}).  We
derived the hardness using {\sl XSPEC} version 12.6
(\citealt{1996adass...5...17A}) by determining the count rates in the
soft and hard band taking the response and ancillary response file for
each of the sources. The three brightest sources suffer from photon
pile-up, where more than one photon is registered during the CCD
integration time (see e.g.~\citealt{2001ApJ...562..575D}). Photon
pile-up will artificially harden those 3 sources in a
hardness--intensity diagram, therefore, we have not plotted these 3
sources in Figure ~\ref{hardnessinten} nor in Table~\ref{srclist}. 
For the other 86 detected sources photon pile--up is unimportant (less than a few percent).
Naively, one would expect most hard sources to be more distant and
more reddened than the soft sources, as the intrinsic spectral shape
of the most numerous classes of sources we expect to find does not
differ much.

The most interesting aspect from Figure~\ref{hardnessinten} is perhaps
the paucity of bright hard sources. Conversely, most bright sources
are probably nearby with low extinction. As foreseen, the spectral
information is insufficient for source classification for the majority
of the total number of detected sources, therefore, classification
will have to come from multi-wavelength observations.  Finally, there
seems to be a dichotomy in the hardness with one peak centered on a
hardness of 0.4 and another centered on -0.5 with a clear paucity of
sources with hardness 0. A similar dichotomy was reported in
\citet{2010arXiv1012.1469W}. 

A conceivable explanation for the nature of this dichotomy is that the
soft sources are dominated by nearby sources, such as RS CVns.  The
harder sources are further away, and are either intrinsically hard
spectrum sources or they appear hard due to the effects of
extinction. The latter increases strongly towards the Galactic center
for distances between $\approx 3-5$~kpc
(\citealt{2006A&A...453..635M}). The strong increase in extinction
over a small distance interval would, in this scenario, be responsible
for the lack of sources near a hardness of 0.

\section{Multi-wavelength observations} 

Optical imaging observations covering the full GBS area have been
obtained using the prime focus MOSAIC~II instrument mounted on the 4~m
Victor M.~Blanco telescope at the Cerro Tololo Inter--American (CTIO)
observatory.  We employed the Sloan \r, \i, and H$\alpha$ filters with
exposures of 120, 180 and 480 seconds, respectively.

For the optical astrometry we used the UCAC2 catalog
(\citealt{2004AJ....127.3043Z}). The UCAC2 is made up by stars
brighter than 16${\rm ^{th}}$ magnitude that would saturate in the
120~s--long \r--band observations. Therefore, to facilitate accurate
astrometric calibration of the optical observations we have also
obtained short, 10~s long, \r--band exposures for each pointing.
A detailed discussion of the optical observations and the findings
will be presented in a forthcoming paper (Bassa et al.~in
preparation).

Using the same telescope and instrument (Blanco and MOSAIC~II), we
re--imaged the fields of the initial optical observations at random
intervals over 9 nights in the Sloan \r--band. The goal of these
observations is to search for variability induced by binary motion. A
detailed discussion of these optical variability observations and the
findings will be presented in a forthcoming paper (Hynes et al.~in
preparation).

In addition to the optical observations, the survey field has also
partially been observed in the near--infrared as part of the UKIRT
Infrared deep sky survey (UKIDSS) Galactic plane survey (UKIDSS/GPS;
\citealt{2008MNRAS.391..136L}).  Furthermore, the GBS area is fully
covered as part of the Vista Variables in the Via Lactea (VVV) survey
(\citealt{2010NewA...15..433M}; Greiss, Steeghs et al.~in prep). These
near--infrared observations will also help us distinguishing between
the different expected source types.  Finally, the Spitzer Galactic
Legacy Infrared Mid-plane Survey Extraordinaire (GLIMPSE;
\citealt{2009PASP..121..213C}) survey has covered the GBS area,
providing mid-infrared imaging of the survey area.

We have also started with the optical spectroscopic identification of
the detected X--ray sources using a suite of optical telescopes such
as the Blanco telescope, the New Technology Telescope, the Magellan
telescopes, the Gemini-South telescope and the Very Large Telescope.
Results of these observations will also be presented in forthcoming
papers.

\section{Comparison with ROSAT all sky survey sources}

In order to investigate whether bright sources in our source list are
detected by the ROSAT All Sky Survey (RASS;
\citealt{1999A&A...349..389V}) and, conversely, to investigate if
bright RASS sources are still detected in our \chan\, observations, we
searched for ROSAT sources within 30\arcsec\,of our \chan\, positions.

\subsection{Persistent sources among \chan\, source \#1--10}

The brightest \chan\, source, source \#1 is a transient (see below).
The four ROSAT Bright Source Catalog (BSC) sources,
1RXS~J173728.0-290759, 1RXS~J174043.1-281806, 1RXS~J173933.4-291001,
1RXS~J173826.7-290140, probably correspond to \chan\, source number 2,
3, 4, and 7 in Table~\ref{srclist}, as the nominal offset between the
\chan\, and ROSAT position is, 5.76, 4.26, 29.7, and 11.6\arcsec,
respectively.  The ROSAT positional (1$\sigma$) uncertainty of these
four sources is 9, 8, 13, and 16\arcsec, respectively. Source \#2 of
the \chan\, GBS survey (1RXS~J173728.0-290759) is a well-known Seyfert
1 galaxy.  Source \#3 (1RXS~J174043.1-281806) is the persistent LMXB
and UCXB-candidate SLX~1737-282. Source \#4 is associated with
HD~316072; the angular distance between the \chan\, position and
HD~316072 is 0.46\arcsec. HD~316072 is a bright, V=9.92, G9III
star. This source is conceivably a bright active star or a long-period
LMXB. Taking the observed and absolute $V$-band magnitude and assuming
${\rm A_V}=0$ yields a maximum source distance of 675 pc. Modelling
the \chan\, X--ray spectrum consisting of 238 photons, we find that
the source spectrum is soft.  It can be described by an absorbed black
body spectrum with kT=0.22 keV and ${\rm
  N_{\rm H}=0.4\times10^{22}\,cm^{-2}}$.  This provides a good fit at an
absorbed/unabsorbed 0.5-10 keV flux of
5.6$\times10^{-13}$/1.2$\times10^{-12}$~\flx, respectively. This
converts to a 0.5-10 keV luminosity upper limit of
6.5$\times10^{31}$\lum. Such a soft spectrum would be consistent both
with a quiescent NS LMXB as well as a white dwarf accretor in a CV.
Optical time series spectroscopy should allow us to distinguish
between an active (binary) star or an X--ray binary scenario.  Source
\#7 is 0.34\arcsec\,away from a bright optical source identified as a
pre-main sequence star by \citet{2006A&A...460..695T}.

Similarly, sources \#5, 6, 8-10 are bright enough that they should
have been detected by the RASS BSC. Source \#5 is 2.78\arcsec\,away
from AX~J1740.1-2847, which is identified as a low-luminosity
high-mass X--ray binary pulsar (\citealt{2010MNRAS.402.2388K}). Source
\#6 is a known Be X--ray binary which is listed in the RASS {\it
  faint} source catalog as 1RXS~J174444.7-271326, indicating some
moderate X--ray variability. Source \#8 is nominally 29.8\arcsec\,away
from AX~J1735.1-2930, which is currently unclassified.  Source \#9 is
0.33\arcsec\, away from HD315997, which is an A5 star in a known
eclipsing binary in a 2.8723~d orbital period
(\citealt{1995A&AS..110..367N}; \citealt{2006IBVS.5674....1O}).  The
V-band magnitude varies between 11.21-11.40 with a secondary eclipse
minimum of 11.24 (\citealt{2006IBVS.5674....1O}). The small size of a
neutron star and black hole implies that an accretion disk should be
present if one wants to explain the amplitude and duration of the
secondary eclipse in an X--ray binary context. Taking the observed and
absolute $V$-band magnitude and assuming ${\rm A_V}=0$ yields a
maximum source distance of 695 pc for the A5 star. This implies a
maximum source 0.5--10 keV luminosity of $\approx 6\times
10^{31}$\lum.  It is also possible that the A5 star is in orbit with a
late type star that is not seen in the optical spectrum but which is
responsible for the X--ray emission. Again, orbital phase resolved
spectroscopic observations are necessary to reveal the nature of this
object. 

Source \#10 is 0.47\arcsec\, away from HD~315992 (F8, V=9.98) and
nominally 26.38\arcsec\, away from 1RXS~J173628.8-291055 which is
listed in the RASS {\it faint} source catalog. In the ASCA Galactic
center survey the source is found to have a 0.7-10 keV flux of
1$\times 10^{-12}$~\flx\,(\citealt{2002ApJS..138...19S}). As for
several of the other sources discussed above, multiple optical spectra
of this source will allow us to investigate if they are in a binary
and, if so, what the nature of the second star is.

To summarize, the \chan\, point sources \#2-10 correspond to: the
Seyfert 1 galaxy 1RXS~J173728.0-290759, SLX~1737-282, HD~316072,
AX~J1740.1-2847, 1RXS~J174444.7-271326, 1RXS~J173826.7-290140,
AX~J1735.1-2930, HD~315997 and HD~315992.

\subsubsection{\chan\, light curves} 
We inspect the \chan\, light curves of source \#1-10.  Source \#1, 3,
and 5 show evidence for flare-like variability.  Fitting the light
curve with a constant gives a $\chi^2$ value of 359, 57, and 43 for
20, 17, and 20 degrees of freedom, respectively.

\begin{figure} \includegraphics[angle=0,width=8cm,clip]{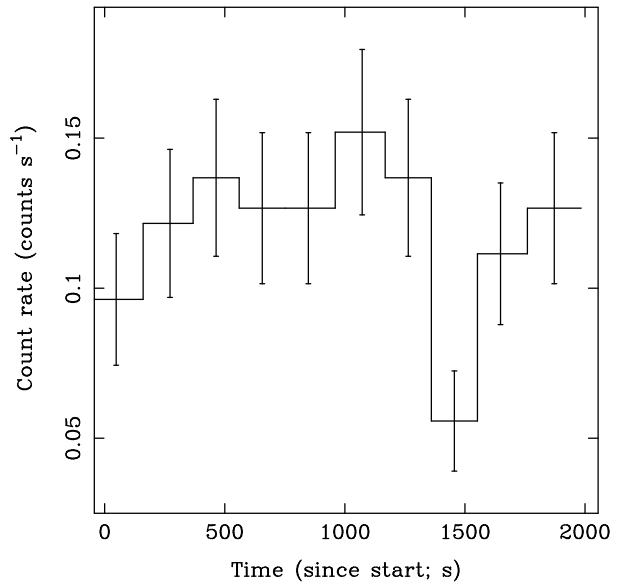}
\caption{The light curve with a 200~s bin size of source \#4 from Table~\ref{srclist}. There is marginal evidence for a dip/eclipse like feature near T=1500~s, although the $\chi^2$ value of a fit of a constant is 27.5 for 20 degrees of freedom.}
\label{lcsrc4} \end{figure}

There is marginal evidence in the light curve of source \#4 for the presence
of an eclipse-like feature (see Figure~\ref{lcsrc4}), although,
fitting the light curve with a constant gives a $\chi^2$ value of 27.5
for 20 degrees of freedom.

The light curves of source \#2, 6, 7, 8, 9, and 10 are consistent with
being constant with $\chi^2$ values of 24.3, 27.2, 22.7, 27.4, 12.4
and 20.5 for 20 degrees of freedom, respectively (except source \#9
for which there are 19 degrees of freedom).

\subsection{Transient sources}
The only other two BSC RASS sources in the GBS area,
1RXS~J175113.2-293842 and 1RXS~J174220.8-273736, are not detected in
our GBS survey. Their nominal ROSAT positional errors are 26\arcsec\,
and 12\arcsec, respectively.  This means that, especially for
1RXS~J175113.2-293842, it is not inconceivable that the source has a
\chan\, counterpart further away than 30\arcsec.  However, even in a
search radius of 1\arcmin\, there is no \chan\, source possibly
related to 1RXS~J175113.2-293842.  Whereas the band passes of ROSAT
(0.1--2.5 keV) and \chan\, (0.3--8 keV) are different, \chan\, is much
more sensitive and both sources should have been exceptionally soft
for it not to be detected in \chan; this is not the case. The ROSAT
hardness ratio 1 (HR1) for 1RXS~J175113.2-293842 is $1.0\pm0.3$ and
hardness ratio 2 (HR2) is $0.54\pm0.27$ (where HR1= (B-A)/(B+A) and
HR2= (D-C)/(D+C), with A=0.11-0.41 keV, B=0.52-2.0 keV, C=0.5-0.9 keV,
and D=0.9-2.0 keV count rate). Assuming a power law spectrum with
index 2, an interstellar extinction of 1$\times 10^{22}$ cm$^{-2}$
(consistent with the observed hardness ratio HR1 and HR2) and
using the RASS-measured 0.1-2.4 keV count rate of 0.07 counts
s$^{-1}$, we calculate the absorbed 0.1-2.4 keV flux to be
approximately $1\times10^{-12}$~\flx\, at the time of the ROSAT
observation. Assuming \chan\, detected zero counts and following
\citet{1986ApJ...303..336G}, we take an upper limit of 3 counts for a
95 per cent upper limit over the 2~ks exposure. This makes the
approximate \chan\, 0.1-2.4 keV upper limit for such a source spectrum
$5\times 10^{-15}$~\flx, implying a decay in flux of at least a factor
200.

For 1RXS~J174220.8-273736 the detected RASS count rate is 0.16 counts
s$^{-1}$, the HR1 is 1.00$\pm$0.03 and the HR2 is 1.00$\pm$0.08. For
the same assumed source spectrum as above the absorbed 0.1-2.4 keV
ROSAT flux was $3\times10^{-12}$~\flx, implying that the source decayed
in flux more than a factor of 500.

Conversely, the brightest source in our survey, SAX~J1750.8-2900
(\citealt{1999ApJ...523L..45N}), was not detected in the RASS BSC.
This source is a recurrent transient LMXB that happened to be in
outburst when our \chan\, observation was obtained but that was
apparently in quiescence during the RASS observations. 

\subsection{Sources from pointed ROSAT observations}
We list ROSAT sources discovered in pointed observations when they
fall inside the covered GBS area. Sources found using both the
position sensitive proportional counter (PSPC) detector as well as
those found using the high resolution imager (HRI) observations are
included (see Table~\ref{pointedrosatlist}). We exclude pointed
observations of sources discussed above (\chan\, source numbers 1-10).

Sometimes multiple ROSAT observations of the same source region
provide slightly different source coordinates, and thus different
ROSAT names, whereas using \chan\, we find one source. Although it is
possible that there was indeed more than one source and these have
faded below the \chan\, detection level, it is more likely that these
multiple ROSAT detections are in fact of one and the same source. As
an example, the six ROSAT sources 2RXP~J173826.2-290147,
2RXP~J173827.4-290138, 2RXP~J173827.5-290152, 2RXP~J173826.7-290205,
2RXP~J173827.2-290144, 2RXP~J173824.0-290146 are probably all related
to \chan\, source \#7 (at $\alpha$=17:38:26.18 and
$\delta$=-29:01:49.4). The nominal angular distance between the ROSAT
positions and the single \chan\, position is 2.45\arcsec, 18.4\arcsec,
16.8\arcsec, 16.7\arcsec, 14.9\arcsec, 28.3\arcsec, respectively. The
last ROSAT position was derived from an observation that had the
source far off-axis (50.8\arcmin). Similarly, the smallest positional
offset was obtained for the observation where the source was only
14.8\arcmin\, off-axis.

\begin{table}
\caption{Sources from pointed ROSAT observations that fall in the GBS area.}
\label{pointedrosatlist}
\begin{center}
\begin{tabular}{llcc}
ROSAT name & \chan& $\alpha$ & $\delta$\\
                        &            &   (hh:mm:ss) & (dd:mm:ss)\\             
\hline
1RXH J173547.0-302851$^{1}$ & \# 43 & 17:35:45.53 & -30:29:00.0 \\
1RXH~J173803.6-290706 & \# 31 & 17:38:03.50 & -29:07:06.1 \\
2RXP~J173940.7-285116 & \# 115 & 17:39:40.80 & -28:51:11.7 \\
2RXP~J174046.7-283849 & \# 469 & 17:40:46.58 & -28:38:50.6 \\
2RXP~J174104.6-281504$^{1}$ & \# 32 & 17:41:04.91 & -28:15:03.4 \\
2RXP~J174133.7-284035$^{1}$ & \# 21 & 17:41:33.76 & -28:40:33.8 \\
2RXP~J174141.9-283324$^{1}$ & \# 114 & 17:41:41.76 & -28:33:24.2 \\
2RXP~J174249.9-275028 & \# 785 & 17:42:49.95 & -27:50:38.5 \\
2RXP~J174834.7-295730 &\#  230$^2$ & 17:48:35.26 & -29:57:31.9 \\
2RXP~J174834.7-295730 &\#  33$^2$ & 17:48:35.54 & -29:57:28.8 \\
2RXP~J174928.4-291901$^{1}$ & \# 156 & 17:49:28.31 & -29:18:59.4 \\
2RXP~J175029.3-285954 & \# 13 & 17:50:29.13 & -29:00:02.3 \\
2RXP~J175041.2-291644$^{1}$ & \# 183 & 17:50:41.17 & -29:16:44.5 \\
\end{tabular}
\end{center}
{\footnotesize $^1$ Multiple sources from the ROSAT pointed source catalog are consistent with the \chan\, position. We list the nearest one.}\\
{\footnotesize $^2$ These two \chan\, sources are 4.8\arcsec\, apart; these are likely two detections of the same \chan\, source.}

\end{table}

\section{Summary}

We have started the \chan\, Galactic Bulge Survey (GBS) with the
goals:\newline i) to identify quiescent, eclipsing NS and BH X--ray
binaries that are bright enough in the optical or near--infrared to
allow phase resolved spectroscopic observations tailored to measure
the compact object mass.  \newline ii) to constrain binary population
synthesis and X--ray binary formation and evolution models by means of
a source number count and a study of the spatial distribution of
X--ray binaries.

In this paper we have presented the \chan\, source list and some
properties of the X--ray sources of observations covering
$\sim$two-thirds ($\approx 8.3$ square degrees) of the total envisaged
survey area of 12 square degrees. The accurate \chan\, source position
will help identify the optical counterparts. The 1234 X--ray sources
that have been discovered so far compares well with the total number
of $\approx 1650$ X--ray sources that we predict we should detect in
the full 12 square degrees. However, this is of course no guarantee
that the number of sources per source class are close to those we
calculated.

We compared our source list with the source list of the RASS. Two BSC
RASS sources, 1RXS~J175113.2-293842 and 1RXS~J174220.8-273736, are not
detected in our GBS survey, indicating a decrease in flux with factors
larger than 200 and 500, respectively. Furthermore, we compared our
\chan\, source list with the sources found in the catalog of sources
derived from pointed HRI and PSPC ROSAT observations that fall inside
the GBS area (see Table 3).

We also imaged the complete survey area using optical observations
obtained with the 4m Blanco telescope at CTIO. Furthermore, we
re-imaged the survey area $\approx$30 times in the $r^\prime$ band to
search for (periodic) variable sources. Results of these campaigns
will be presented in forthcoming papers.

\section{Outlook}

We are in the process of obtaining optical spectroscopic and
photometric observations of all optical proposed counterparts in the
\chan\, error circle to classify the X--ray sources. This step is
crucial to achieve our science goals outlined above.

A full discussion of all known information on all the X--ray sources is
beyond the scope of this paper and we defer the discussion of the
other fainter \chan\, sources to a forthcoming paper, where they will
be discussed together with optical photometric and spectroscopic
information.

\section*{Acknowledgments} \noindent P.G.J. and G.N. acknowledge support
from the Netherlands Organisation for Scientific Research. R.I.H.,
C.T.B, V.J.M., and L.G. would like to acknowledge support from
National Science Foundation Grant No.\ AST-0908789 and NASA/Louisiana
Board of Regents grant NNX07AT62A/LEQSF(2007-10) Phase3-02. COH is
supported by NSERC. DS acknowledges an STFC Advanced Fellowship. TJM
and AD thank STFC for support via a rolling grant to the University of
Southampton.

\end{document}